\documentclass[longauth,letter]{aa}
\usepackage{graphicx}
\usepackage{txfonts}
\usepackage{natbib}
\bibpunct{(}{)}{,}{a}{}{,}
\begin{document}
\newcommand{\doceCO}{\mbox{$^{12}$CO}}
\newcommand{\doce}{\mbox{$^{12}$CO}}
\newcommand{\trece}{\mbox{$^{13}$CO}}
\newcommand{\treceCO}{\mbox{$^{13}$CO}}
\newcommand{\CdieciochoO}{C\mbox{$^{18}$O}}
\newcommand{\jdn}{\mbox{$J$=10$-$9}}
\newcommand{\jdsq}{\mbox{$J$=16$-$15}}
\newcommand{\jsc}{\mbox{$J$=6$-$5}}
\newcommand{\jtd}{\mbox{$J$=3$-$2}}
\newcommand{\jcc}{\mbox{$J$=5$-$4}}
\newcommand{\jdu}{\mbox{$J$=2$-$1}}
\newcommand{\juc}{\mbox{$J$=1$-$0}}
\newcommand{\gsim}{\raisebox{-.4ex}{$\stackrel{>}{\scriptstyle \sim}$}}
\newcommand{\lsim}{\raisebox{-.4ex}{$\stackrel{<}{\scriptstyle \sim}$}}
\newcommand{\psim}{\raisebox{-.4ex}{$\stackrel{\propto}{\scriptstyle \sim}$}}
\newcommand{\kms}{\mbox{km~s$^{-1}$}}
\newcommand{\s}{\mbox{$''$}}
\newcommand{\mloss}{\mbox{$\dot{M}$}}
\newcommand{\my}{\mbox{$M_{\odot}$~yr$^{-1}$}}
\newcommand{\ls}{\mbox{$L_{\odot}$}}
\newcommand{\ms}{\mbox{$M_{\odot}$}}
\newcommand{\mm}{\mbox{$\mu$m}}
\def\arcdeg{\hbox{$^\circ$}}
\newcommand{\seca}{\mbox{\rlap{.}$''$}}
\newcommand{\dega}{\mbox{\rlap{.}$^\circ$}}
\newcommand{\aprop}{\raisebox{-.4ex}{$\stackrel{\propto}{\scriptstyle\sf \sim}$}}
\newcommand{\apropg}{\raisebox{-.4ex}{$\stackrel{\Large \propto}{\sim}$}}
\title{Herschel/HIFI observations of high-$J$ CO transitions in the
  protoplanetary nebula CRL 618\thanks{Herschel is an ESA space
  observatory with science instruments provided by European-led
  Principal Investigator consortia and with important participation
  from NASA.}}

   \subtitle{}

   \author{V. Bujarrabal
          \inst{1}
          \and
J. Alcolea
          \inst{2}
\and
R. Soria-Ruiz
\inst{2}
\and
P. Planesas
\inst{1,12}
\and
D. Teyssier
          \inst{7}
\and
A.P. Marston
\inst{7}
\and
J. Cernicharo
\inst{3}
\and
L. Decin
\inst{4,5}
\and
C. Dominik
\inst{5,14}
\and
K. Justtanont
\inst{6}
\and
A. de Koter
\inst{5,15}
\and
G. Melnick
\inst{8}
\and
K.M. Menten
\inst{9}
\and
D.A. Neufeld
\inst{10}
\and
H. Olofsson
\inst{6,11}
     \and 
M. Schmidt
\inst{13}
\and
F.L. Sch\"oier
\inst{6}
     \and 
R. Szczerba
\inst{13}
\and
L.B.F.M. Waters
\inst{5,4}
\and
G. Quintana-Lacaci
\inst{16}
R. G\"usten
\inst{9}
\and
J.D. Gallego
\inst{17}
\and
M.C. D\'{\i}ez-Gonz\'alez
\inst{17}
\and
A. Barcia
\inst{17}
\and
I. L\'opez-Fern\'andez
\inst{17}
\and
K. Wildeman   
\inst{18}      
\and
A.G.G.M. Tielens  
\inst{19}
\and
K. Jacobs      
\inst{20}
          }


   \institute{   
Observatorio Astron\'omico Nacional (IGN), Ap 112, E--28803 
Alcal\'a de Henares, Spain\\
              \email{v.bujarrabal@oan.es}
	      \and
Observatorio Astron\'omico Nacional (IGN), Alfonso XII N$^{\circ}$3,
              E--28014 Madrid, Spain  
\and
CAB, INTA-CSIC, 
Ctra de Torrej\'on a Ajalvir, km 4,
E--28850 Torrej\'on de Ardoz, Madrid, Spain
\and
Instituut voor Sterrenkunde,
             Katholieke Universiteit Leuven, Celestijnenlaan 200D, 3001
Leuven, Belgium
\and
    Sterrenkundig Instituut Anton Pannekoek, University of Amsterdam,
Science Park 904, NL-1098 Amsterdam, The Netherlands
\and
Onsala Space Obseravtory,  Dept. of Radio and Space Science, Chalmers  
University of Technology, SE--43992 Onsala, Sweden
\and
European Space Astronomy Centre, ESA, P.O. Box 78, E--28691
Villanueva de la Ca\~nada, Madrid, Spain
\and
Harvard-Smithsonian Center for Astrophysics, Cambridge, MA 02138, USA
\and
Max-Planck-Institut f{\"u}r Radioastronomie, Auf dem H{\"u}gel 69,
D-53121 Bonn, Germany 
\and
The Johns Hopkins University, 3400 North Charles St, Baltimore, MD  
21218, USA
\and
Department of Astronomy, AlbaNova University Center, Stockholm  
University, SE--10691 Stockholm, Sweden
\and
Joint ALMA Observatory, El Golf 40, Las Condes, Santiago, Chile
\and
N. Copernicus Astronomical Center, Rabia{\'n}ska 8, 87-100 Toru{\'n}, Poland
\and 
Department of Astrophysics/IMAPP, Radboud University Nijmegen,   
Nijmegen, The Netherlands
\and
Astronomical Institute, Utrecht University,
Princetonplein 5, 3584 CC Utrecht, The Netherlands 
\and
Instituto de Radioastronom{\'i}a Milim\'etrica (IRAM),
Avda.\ Divina Pastora 7, E--18012 Granada, Spain
\and
 Observatorio Astron\'omico Nacional (IGN), 
Centro 
Astron\'omico de Yebes, Apartado 148, E--19080 Guadalajara, Spain
\and
SRON, Netherlands Institute for Space Research, 
Landleven 
12, 9747 AD Groningen, The Netherlands
\and
             Sterrewacht Leiden, University of Leiden, P.O. Box 9513,
	     2300 RA Leiden, The Netherlands 
\and
              KOSMA, I. Physik. Institut, Universit\"at zu 
K\"oln, Z\"ulpicher 
Str. 77, D 50937 K\"oln,   Germany
}

   \date{Received ... ; accepted ...}

 
  \abstract
   {}
   {We aim to study the physical conditions, particularly the
  excitation state, of the intermediate-temperature gas components
  in the protoplanetary nebula CRL\,618. These components 
  are particularly important for understanding the evolution of the nebula.}
   {We performed Herschel/HIFI observations of several CO  
  lines in the far-infrared/sub-mm in the protoplanetary nebula CRL\,618. 
The high spectral resolution provided by HIFI
  allows measurement of the line profiles. Since the
  dynamics and structure of the nebula is well known from  
  mm-wave interferometric maps, it is possible to identify the
  contributions of the 
  different nebular components (fast bipolar outflows, double shells, compact
  slow shell) to the line profiles. The observation of
  these relatively high-energy transitions allows an accurate study of the
  excitation conditions in 
these components, particularly in the warm ones, which
  cannot be properly studied from the low-energy lines.
  }
  {The \doce\ $J$=16$-$15, 10$-$9, and 6$-$5 lines are easily 
  detected in this source. \trece\ $J$=10$-$9 and 6$-$5 are
  also detected. Wide profiles showing spectacular line wings have been
  found, particularly in \doce\ \jdsq. Other lines observed
  simultaneously with CO are also shown. Our analysis of the CO
  high-$J$ transitions, when compared with the existing models,
  confirms the very low expansion velocity of the central, dense
  component, which probably indicates that the shells ejected during
  the last AGB phases were driven by radiation pressure under a regime
  of maximum transfer of momentum. No contribution of the diffuse halo
  found from mm-wave data is identified in our spectra, because of its
  low temperature.  
  We find that the fast bipolar outflow is quite hot, much hotter than 
  previously estimated; for instance, gas flowing at 100
  \kms\ must have a temperature higher than $\sim$ 200
  K. Probably, this very fast outflow, with a kinematic age $<$ 100 yr,
  has been accelerated by a shock and has not yet cooled down.
  The
  double empty shell found from mm-wave mapping must also be
  relatively hot, in agreement with the
  previous estimate.
}
{}


   \maketitle
%

\section{Introduction}

Protoplanetary nebulae (PPNe) are known to present very fast bipolar
outflows, along with slower components, which are probably the remnants
of the mass ejection during the previous AGB phase. The bipolar flows
typically reach velocities of 100 \kms, and affect a sizable fraction
of the nebular mass, $\sim$ 0.1 -- 0.3 \ms\ \citep{bujetal01}. These
dense flows actually represent intermediate states in the spectacular
evolution from the spherical and slowly expanding circumstellar
envelopes around AGB stars to the planetary nebulae, which usually show
bipolar or ring-like symmetries. Such remarkable dynamics is probably
the result of the interaction between the AGB and post-AGB winds:
axial, very fast post-AGB jets colliding with the denser material
driven isotropically away from the star during its AGB phase
\citep[e.g.][]{balickf02}. The presently observed bipolar
outflows would then correspond to a part of the relatively dense shells
ejected during the last AGB phase, mostly their polar regions,
accelerated by the shocks that propagate during the PPN phase.

The massive bipolar outflows in PPNe, as well as the unaltered remnants
of the AGB shells, usually show strong emission in molecular lines
\citep{bujetal01}. PPNe have been accurately observed in mm-wave lines,
particularly by means of interferometric maps with resolutions $\sim$
1$''$. Thanks to those observations, the structure, dynamics, and
physical conditions in these nebulae are often quite well
known. However, observations of the low-$J$ transitions are not very
useful for studying the warm gas components. The well-studied \jdu\ and
\juc\ transitions of \doce\ only require temperatures of $T_{\rm k}$
$\sim$ 15 K to be excited.
Indeed their maximum emissivity
occurs for excitation temperatures of 10 -- 20 K, and the line
intensities and line intensity ratios only slightly depend on the
excitation state in relatively warm gas. Needless to
say, observations in the visible or near infrared ranges tend to select
hot regions, with typical temperatures over 1000 K. The proper study of
warm regions, 100 K \lsim\ $T_{\rm k}$ \lsim\ 1000 K, therefore requires
observations at intermediate wavelengths, in the far infrared (FIR) and
sub-mm ranges.

Because of the role of shocks in PPN evolution, these warm regions are
particularly important for understanding nebular structure and
evolution. In some well-studied cases, e.g.\ M\,1--92 and M\,2--56
\citep{alcolea07,bujetal98,ccarrizo02}, the high-velocity, massive
outflows are found to be very cold, with temperatures \lsim\ 20 K,
which implies very fast cooling in the shock-accelerated gas. No
warm component representing the gas recently accelerated by the shock
front has been identified in these sources.  In other cases such
as CRL\,618,
interferometric imaging of the \doce\ \jdu\ line shows that dense gas
in axial structures presents higher temperatures \citep[][hereafter,
SC04]{sanchezc04}.  But, precisely because of their relatively high
excitation, the temperature estimate in these components from CO
\jdu\ is very uncertain.  Indeed, even the presence of such high
excitation in the dense bipolar outflows in CRL\,618 remained to be
demonstrated.

Other attempts to study the warm gas in CRL\,618 were carried out by
({\em i}) \cite{justtanont00} from ISO data with low spectral resolution;
({\em ii}) \cite{pardo04}, who focused on the chemistry of the different
components; and ({\em iii}) \cite{naka07}, who also obtained maps of the
\jsc\ transition in CRL\,618, but with less detail than in SC04.

The Herschel Space Telescope is well-suited to studying warm gas
around evolved stars in the FIR and sub-mm. The high spectral
resolution that can be achieved with its heterodyne instrument HIFI
(better than 1\,\kms) is particularly useful for this purpose, since
kinematics offers a fundamental key to understanding this warm, shocked
gas. Here we present Herschel/HIFI observations of CRL\,618 in several
molecular lines of \doce\ and \trece\ that were obtained as part of the
guaranteed-time key program HIFISTARS, which is devoted to the study of
intermediate-excitation molecular lines in nebulae around evolved
stars.

\begin{figure}
\rotatebox{270}{\resizebox{12cm}{!}{ 
\includegraphics{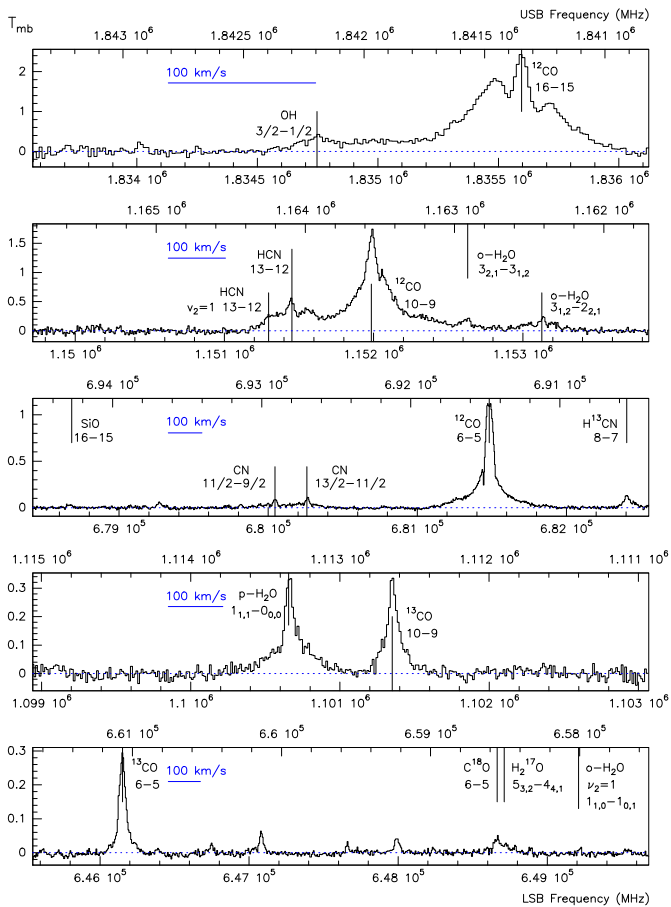}
}}
\caption{HIFI observations containing detected \doce\ and \trece\ lines
  in CRL\,618 ($T_{\rm mb}$ vs.\ rest frequencies).
  Frequency scales and detected
  lines are indicated; note that the
  observations are performed in DSB mode.} 
\end{figure}

\section{Observations}

We used the Herschel/HIFI instrument \citep{pilbratt10, degraauw10} to
observe the \jsc, 10$-$9, and 16$-$15 transitions of \doceCO\ and
\treceCO\ in the PPN CRL\,618 (\treceCO\ \jdsq\ was not detected); see
Figs.\ 1 and 2.  Other molecular lines were also detected within the
observed frequency ranges.  The data were taken using the two
orthogonal HIFI receivers available at each band. Both receivers
work in double side-band (DSB) mode, which effectively doubles the
instantaneous IF coverage. Care was taken when choosing the local
oscillator frequency, maximizing the number of observed interesting
lines.

The observations were obtained in the dual-beam-switching (DBS)
mode. In this mode, the HIFI internal steering mirror chops between the
source position and an emission-free position 3$'$ away. The
telescope then alternatively locates the source in either of the
chopped beams, providing a double-difference calibration scheme, which
allows a more efficient cancellation of the residual standing waves in
the spectra (see additional details in
Helmich et al.\ 2010). This procedure works very well
except for the \jdsq\ lines, where strong ripples were found in some
spectra, especially in the V-receiver. 

The HIFI data shown here were taken using the Wide-Band
Spectrometer (WBS), an acousto-optical spectrometer that provides
simultaneous coverage of the full instantaneous IF band in the two
available orthogonal receivers, with a spectral resolution of 1.1 MHz. 
The data shown in all the
figures have been resampled and smoothed to a resolution of about
2\,\kms.

\begin{table}
\caption{Summary of the Herschel telescope characteristics.}
\begin{center}
\begin{tabular}{lccccc}
line & band & DSB $T_{\rm sys} $ & HPBW & $T_{\rm mb}/T_{\rm a}$ &
cal.\ uncert.\ \\ 
\hline
6--5      & 2 & 130 K  & 31$''$ & 1.4 & 10\% \\
\trece\ 10--9 & 4 & 400 K  & 20$''$ & 1.4 & 20\%  \\
\doce\ 10--9  & 5 & 800 K  & 20$''$ & 1.4 & 15\% \\
16--15        & 7 & 1300 K & 12$''$ & 1.5 & 30\% \\
\hline
\end{tabular}
\end{center}
\end{table}

The data were processed with the standard HIFI pipeline using HIPE,
with a modified version of the level 2 algorithm that yields unaveraged
spectra with all spectrometer sub-bands stitched together.  Later on,
the spectra were exported to CLASS using the hiClass tool within HIPE,
for further inspection, flagging out data with outstanding ripple
residuals, final averaging, and baseline removal. We checked that the
profile of the \doce\ \jdsq\ line, in particular, is not significantly
affected by ripples, which in fact are not noticeable even in the
relatively flat parts of the final spectrum.  The data were originally
calibrated in antenna temperature units and later converted into
main-beam temperatures ($T_ {\rm mb}$).  In all cases we assumed a 
side-band gain ratio of one.  A summary of the telescope
characteristics and observational uncertainties is given in Table 1.

We also report here \doceCO\ \mbox{$J$=4$-$3} and 7$-$6
lines observed from the ground with the APEX telescope. These
observations were performed in preparation for the HIFI observations in
2006 and more details will be given in a future paper. We used the FLASH
receiver equipped with two FTS spectrometers, which allows 
simultaneous observation of the two CO lines; see \cite{heyminck06} for
a description of the system.  The resulting APEX spectra are shown
in Fig.\,2, after being rescaled to $T_ {\rm mb}$ units using the
values for the telescope efficiencies given by \cite{gusten06}. In this
figure, we also show IRAM 30m data data for \doceCO\ \juc\ and 2$-$1
from \cite{bujetal01}.

\section{Nebula model}

Detailed mapping of CO emission at 1mm
wavelength was performed by SC04, who derived
the physical conditions, structure, and dynamics of the nebula from
model fitting of their maps. Several warm components were identified
(see Fig.\ 3): a
compact dense core with temperatures $\sim$ 100 K (decreasing with
distance to the center), a double (empty) shell with a typical
temperature of 200 K, and a very fast bipolar outflow (running inside
the cavities) with temperatures \lsim\ 100 K (also decreasing with
distance). Another cooler component was found: a diffuser
halo expanding at 17 \kms. The empty shells show a roughly elliptical
shape that strongly suggests that they are the result of a bow-like shock,
but with a moderate expansion velocity. The
expansion velocity of the compact component was found to be
particularly low, \lsim\ 10 \kms.

As a starting point, we adopted a similar description of the nebula
structure. Calculation of the expected emission in our case must be
more sophisticated, because the high-$J$ CO transitions cannot be
assumed to be thermalized in the whole nebula. Therefore, we determined
the CO level population performing LVG calculations for a large number
of points in the model nebula. LVG calculations are fully justified
here because of the high velocity gradients characteristic of this
source. With these values of the populations we calculated the
emissivity and absorption coefficients at each point. Finally, the
brightness distribution is calculated along a number of lines of sight
and is convolved with the telescope beam shape, described by a Gaussian
function. The results are profiles in units of main-beam temperature,
directly comparable to our observations. 
Therefore, opacity effects are properly   
taken into account, both in the excitation and line
profile calculations. We have checked that the
high-$J$ lines are often underexcited in our case, particularly for
densities under $\sim$ 10$^6$ cm$^{-3}$. 

More detailed discussions of the code and involved uncertainties will
be presented in a forthcoming paper. Here we mostly discuss a
preliminary comparison of the model results with our observations of
the \doce\ $J$=16$-$15 line. The model fitting cannot be considered as
satisfactory without studying all the observed profiles, as well as the
mm-wave maps, requiring a very careful treatment of the numerical
uncertainties and a detailed discussion of the nebula structure.

\section{Results}

\begin{figure}
\center{\rotatebox{270}{\resizebox{10cm}{!}{ 
\includegraphics{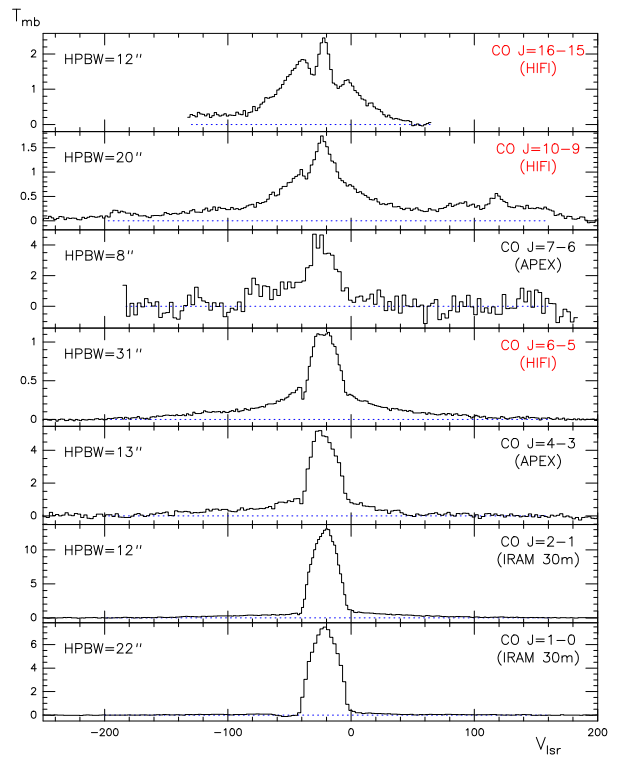}
}}
}
\caption{\doce\ lines observed in CRL\,618 with HIFI, in units of
  $T_{\rm mb}$ vs.\ LSR velocity. Other CO lines observed from the
  ground are also shown.  Note that the high-velocity wings of the
  \doce\ \jdn\ profile are affected by emission of other lines (Fig.\
  1).  }
\end{figure}

We present in Fig.\ 2 our observational results for the \doce\ lines
obtained with HIFI, together with other lines observed from the ground
(Sect.\ 2). The baseline of the \jdsq\ line is not well
determined, because of the lack of spectral coverage and the presence
of other lines, which mostly affects the
intensity of the line wings at extreme velocities. 

We have mentioned (Sect.\ 1) the interest of observing high-$J$
transitions in order to better estimate the excitation conditions in
warm regions. Our highest transitions require
several hundred K to be significantly excited; for instance, the energy
of the 
\doce\ $J$=16 level is equivalent to 750 K.

We have compared our data with the predictions of the model presented
in Sect.\ 3, originally developed to explain the mm-wave maps (see
assumed nebula structure in Fig.\ 3).  This model can reasonably explain
the central part of the high-$J$ emission with only moderate changes.
We can see in Fig.\ 4 the comparison between the observations and
predictions for a model in which the density and temperature of the
shells have been increased
by 25\%\
 (dashed, red line).
The asymmetry in the observed profile cannot be explained by
radiative transfer phenomena, and it probably reveals true
asymmetries with respect to the equator that are
not included in our models. The central spectral feature is very narrow
and comes from the low-velocity compact component of the nebula; in
fact, the fitting is improved if the velocity of central component is
still decreased, to typical values \lsim\ 5 \kms, but keeping the
strong velocity gradient characteristic of this region, in order to
reproduce the triangular shape of the observed feature. The two intense
humps at both sides of the central maximum would come from the hollow
shells.

However, the high-velocity wings are severely underestimated by our
standard model, and we would have to significantly increase the
excitation conditions of the very fast bipolar outflow of CRL\,618 to
reproduce their intensity.  We can also see in Fig.\ 4 our predictions
for a model similar to the previous one, but with $T_{\rm k}$ $\sim$
200 K in the regions that present expansion velocities $\sim$ 100
\kms. Other parameters of the fast outflow, particularly its velocity
and density distributions do not change with respect to the original
model. The asymmetry between the red and blue line wings is not
reproduced by our predictions also in this case. It is remarkable that
the high-velocity outflow obviously contributes to the emission at the
profile central features. Therefore, in this model the requirements to
reproduce the secondary maxima are significantly weaker, and a
temperature similar to or even lower than assumed in our original model
for the empty shells would be compatible with the observations. The
general properties of the central, dense component mentioned before, in
particular its low velocity, are required in this case too. The high
velocity wings are also detected in emission from other molecules, such
as H$_2$O, HCN, and CN (Fig.\ 1), which must be significantly abundant
in this recently shocked gas.

The relatively high temperatures deduced here for the fast bipolar flow
relax the discrepancy usually found between the high excitation of the
shocked gas predicted by theoretical models and the observational results
\citep[an intricate theoretical problem not discussed in this
letter, see e.g.][]{lee09}. 
However, the discrepancy persists. \cite{lee09} predict 
temperatures of the high-velocity gas in CRL\,618 over $\sim$ 1000 K
and too weak CO emission in all rotational lines, since shocks are
expected to dissociate molecules.

The low-excitation component of the nebula model by SC04, the extended
halo, has apparently no counterpart in the observations, because the
model predicts a very low intensity from such cool gas and the high-$J$
line profiles do not seem to require any contribution from it.

\begin{figure}
\hspace{0.6cm}
\rotatebox{270}{\resizebox{7cm}{!}{ 
\includegraphics{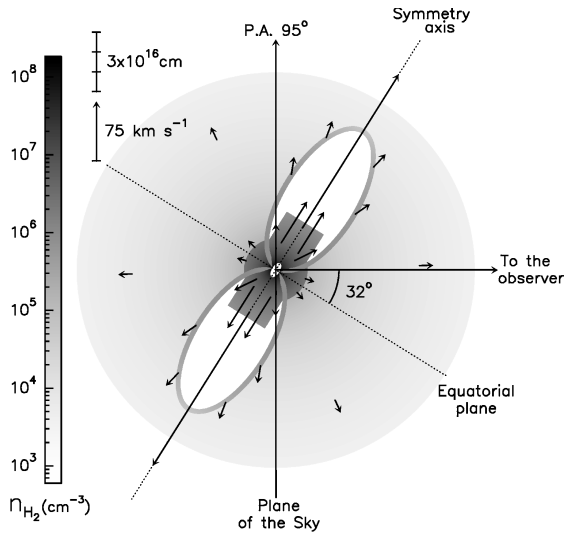}
}}
\caption{Basic model properties used in our calculations, see Sects.\ 3 and
  4 (adapted from SC04).}
\end{figure}

\begin{figure}
\vspace{-1.7cm}
\hspace{-0.6cm}
\rotatebox{270}{\resizebox{7cm}{!}{ 
\includegraphics{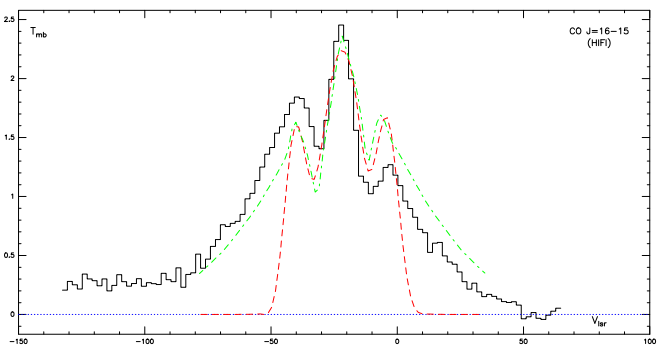}
}}
\vspace{-0.4cm}
\caption{\doce\ \jdsq\ observed line and predictions of our model
  (Sects. 3, 4), without (dashed, red line), and with (dot-dashed, green
  line) significant increase of the
  excitation in the very fast outflow.}
\end{figure}

In Fig.\ 1 we also show our HIFI spectra of \trece\ lines. The \trece\
\jdsq\ observations are not very sensitive, so the line is not detected
with a limit $T_{\rm mb}$ \lsim\ 0.2 K.  As we can see, the contrast
between \doce\ and \trece\ lines is high (mainly in the line wings,
about a factor ten for the highest transitions). This result is
compatible with our calculations, which suggest moderate opacities in
high-$J$ \doce\ lines from the main nebular components, in particular
with $\tau$(16$-$15) \lsim\ 1 for gas flowing at \gsim\ 100 \kms.  

Our results can therefore be summarized as follows

\noindent
1. We detected high-$J$ CO emission using
   Herschel/HIFI. The high-velocity line wings characteristic of this
   source become progressively dominant as the level energies increase,
   with \doce\ \jdsq\ showing a spectacular composite profile.

\noindent
2. The temperature of the very fast bipolar outflow in CRL\,618 is
   high, significantly higher than the previously adopted values. SC04
   proposed a temperature for this component $<$ 100 K, which, in view
   of the intense line wings seen in the \jdsq\ transition, must
   be significantly increased. From our calculations, we estimate that
   gas flowing at about 100 \kms\ must have a temperature $\sim$ 200
   K. We suggest that this very fast outflow, with a kinematic age $<$
   100 yr, was accelerated by a shock and has not yet fully cooled
   down. The rest of the physical conditions are not significantly
   changed, therefore the dynamical parameters (including the high
   momentum and kinetic energy) remain the same as those deduced by
   SC04.

\noindent
3. We confirm the low expansion velocity of the very dense, central
   component. This low velocity coincides with a significant increase
   in the mass-loss rates during the last $\sim$ 400 yr of the AGB
   phase (SC04). We suggest that the significant decrease in the
   expansion velocity is caused by the ejection of material by the star
   during the last AGB phases being driven by radiation pressure under
   a regime of maximum momentum transfer from radiation to gas.

We recall that our analysis is preliminary. Further developments, in
particular including fitting the mm-wave maps and our HIFI data,
are required.

\begin{acknowledgements}
HIFI has been designed and built by a consortium of institutes and
university departments from across Europe, Canada, and the United
States under the leadership of SRON Netherlands Institute for Space
Research, Groningen, The Netherlands, and with major contributions from
Germany, France, and the US.  Consortium members are: Canada: CSA,
U.Waterloo; France: CESR, LAB, LERMA, IRAM; Germany: KOSMA, MPIfR, MPS;
Ireland, NUI Maynooth; Italy: ASI, IFSI-INAF, Osservatorio Astrofisico
di Arcetri- INAF; Netherlands: SRON, TUD; Poland: CAMK, CBK; Spain:
Observatorio Astron\'omico Nacional (IGN), Centro de Astrobiolog\'{\i}a
(CSIC-INTA). Sweden: Chalmers University of Technology - MC2, RSS \&
GARD; Onsala Space Observatory; Swedish National Space Board, Stockholm
University - Stockholm Observatory; Switzerland: ETH Zurich, FHNW; USA:
Caltech, JPL, NHSC. HCSS / HSpot / HIPE is a joint development (are
joint developments) by the Herschel Science Ground Segment Consortium,
consisting of ESA, the NASA Herschel Science Center, and the HIFI,
PACS, and SPIRE consortia. This work has been partially supported by
the Spanish MICINN, program CONSOLIDER INGENIO 2010, grant ``ASTROMOL"
(CSD2009-00038). R.Sz.\ and M.Sch.\ acknowledge support from grant N
203 393334 from the Polish MNiSW. K.J.\ acknowledges the funding from
SNSB. J.C.\ acknowledges funding from MICINN, grant AYA2009-07304. This
research was performed, in part, through a JPL contract funded by the
National Aeronautics and Space Administration.
\end{acknowledgements}

\bibliographystyle{aa}

\begin{thebibliography}{}


\bibitem[Alcolea et al.(2007)]{alcolea07} Alcolea, J., Neri, R., \&
Bujarrabal, V.\ 2007, \aap, 468, L41


\bibitem[Balick \& Frank(2002)]{balickf02} Balick, B., \& Frank, A.\ 2002,
  \araa, 40, 439 

\bibitem[Bujarrabal et al.(1992)]{bujetal92} Bujarrabal, V., Alcolea,
J., \& Planesas, P.\ 1992, \aap, 257, 701

\bibitem[Bujarrabal et al.(1998)]{bujetal98} Bujarrabal, V., 
Alcolea, J., \& Neri, R.\ 1998, \apj, 504, 915 

\bibitem[Bujarrabal et al.(2001)]{bujetal01} Bujarrabal, V.,
Castro-Carrizo, A., Alcolea, J., \& S{\'a}nchez Contreras, C.\ 2001,
\aap, 377, 868  

\bibitem[Castro-Carrizo et al.(2002)]{ccarrizo02} Castro-Carrizo, A.,
Bujarrabal, V., S{\'a}nchez Contreras, C., Alcolea, J., \& Neri, R.\
2002, \aap, 386, 633

\bibitem[de Graauw et al.(2010)]{degraauw10} de Graauw, Th., et
  al.\ 2010, \aap, in press

\bibitem[G{\"u}sten et al.(2006)]{gusten06} 
G{\"u}sten, R., Nyman, L.~{\AA}., Schilke, P., 
Menten, K., Cesarsky, C., \& Booth, 
R.\ 2006, \aap, 454, L13 

\bibitem[Helmich et al.(2010)]{helmich10} Helmich, F., et
  al.\ 2010, \aap, this volume 

\bibitem[Heyminck et al.(2006)]{heyminck06} 
Heyminck, S., Kasemann, C., G{\"u}sten, R., de Lange, 
G., \& Graf, U.~U.\ 2006, \aap, 454, L21 

\bibitem[Justtanont et al.(2000)]{justtanont00} Justtanont, K., et al.\
  2000, \aap, 360, 1117 
 
\bibitem[Lee et al.(2009)]{lee09} Lee, C.-F., Hsu, M.-C., 
\& Sahai, R.\ 2009, \apj, 696, 1630 

\bibitem[Nakashima et al.(2007)]{naka07} Nakashima, J.-i., et 
al.\ 2007, \aj, 134, 2035 

\bibitem[Pardo et al.(2004)]{pardo04} Pardo, J.~R., Cernicharo, 
J., Goicoechea, J.~R., \& Phillips, T.~G.\ 2004, \apj, 615, 495 

\bibitem[Pilbratt et al.(2010)]{pilbratt10} Pilbratt, P., et al.\ 2010,
  \aap, in press

\bibitem[S{\'a}nchez Contreras et al.(2004)]{sanchezc04} 
S{\'a}nchez Contreras, C., Bujarrabal, V., Castro-Carrizo, A., Alcolea, J., 
\& Sargent, A.\ 2004, \apj, 617, 1142 (SC04)


\end{thebibliography}

\end{document}